\documentclass[aps,pra,preprint,a4paper,showpacs,amsmath,amssymb]{revtex4-1}

\usepackage[utf8]{inputenc}

\usepackage{pdfsync}

\usepackage{amsmath}
\usepackage{amssymb}
\usepackage{latexsym}
\usepackage{mathrsfs}

\usepackage{boxedminipage}

\usepackage{ifpdf}

\ifpdf
\usepackage[pdftex]{graphicx}
\else
\usepackage{graphicx}
\fi

\ifpdf
\DeclareGraphicsExtensions{.pdf, .jpg, .tif}
\else
\DeclareGraphicsExtensions{.eps, .jpg}
\fi

\usepackage{float}

\begin{document}

\title{The ground state of the polar alkali-Strontium molecules: potential energy curve and permanent dipole moment}
\author{R. Gu\'erout$^{1}$, M. Aymar$^{2}$ and O. Dulieu$^{2}$}
\affiliation{$^{1}$Laboratoire Kastler-Brossel, CNRS, ENS, Univ Pierre et Marie Curie case 74,
Campus Jussieu, F-75252 Paris Cedex 05, France\\
$^{2}$Laboratoire Aim\'e Cotton, CNRS, UPR3321, B\^at. 505, Univ Paris-Sud, 91405 Orsay Cedex, France}
\email[O. Dulieu: ]{olivier.dulieu@u-psud.fr}

\date{\today}

\begin{abstract}
In this study, we investigate the structure of the polar alkali-Strontium diatomic molecules as possible candidates for the realization of samples of new species of ultracold polar molecules. Using a quantum chemistry approach based on Effective Core Potentials and Core Polarization Potentials, we model these systems as effective three valence electron systems, allowing for calculation of electronic properties with Full Configuration Interaction. The potential curve and the permanent dipole moment of the $^2\Sigma^+$ ground state are determined as functions of the internuclear distances for LiSr, NaSr, KSr, RbSr, and CsSr molecules. These molecules are found to exhibit a significant permanent dipole moment, though smaller than those of the alkali-Rb molecules.
\end{abstract}

\maketitle

\section{Introduction} 
\label{sec:introduction}

The prospects of the realization of quantum degenerate gases composed of polar systems, \textit{i.e.} exhibiting a permanent dipole moment, either magnetic or electric, are among the most fascinating ones in the context of researches on ultracold matter \cite{baranov2007}, related to many-body physics, quantum information, or quantum simulators of solid state physics. Indeed, neutral particles usually weakly interact via van der Waals short-range potentials varying with their mutual distance $R$ as $R^{-6}$, which has been a key condition for the observation of Bose-Einstein condensation in weakly interacting gases of ultracold alkali atoms \cite{anderson1995,davis1995,bradley1995}. In contrast, particles with a permanent dipole moment interact through a long-range dipole-dipole potential varying as $R^{-3}$ which dominates the van der Waals interaction. Ultracold gases of such particles are predicted to evolve in a entirely new regime of strong interactions with pronounced anisotropy. For instance, the Chromium atom has a large magnetic dipole moment of about $6\mu_B$, and once condensed \cite{griesmaier2005,beaufils2008}, the manifestation of anisotropic interactions has been observed experimentally \cite{stuhler2005,beaufils2009}. Ground state polar molecules, having permanent dipole moment (in the body-fixed frame) ranging between a few tenths of a Debye like LiNa or KRb, up to several Debye like LiCs \cite{aymar2005} are actually strongly interacting with larger forces than in the Chromium case, and are expected to be better candidates for studying strongly interacting degenerate gases. Up to now, several attempts to observe ultracold ground state polar molecules have been successful with bialkali molecules, namely KRb \cite{mancini2004,wang2004}, LiCs \cite{deiglmayr2008}, NaCs \cite{haimberger2004}, RbCs \cite{kerman2004}. However the achievement of quantum degeneracy of ultracold molecules is conditioned by the phase space density of molecules in a unique quantum state, preferably the absolute ground state, defined not only by their vibrational and rotational quantum numbers, but also by their hyperfine structure \cite{ni2008,ospelkaus2010b,danzl2010}.

As the number of atomic species other than alkalis, which are laser-cooled and subsequently brought to quantum degeneracy is continuously increasing, the formation of novel species of polar molecules will soon be investigated in the ultracold regime \cite{meyer2009}. For instance, photoassociation (PA) of Yb$_2$ \cite{takasu2004,kitagawa2008} and YbRb \cite{nemitz2010} molecules has already been reported, while quantum degeneracy in Ytterbium \cite{fukuhara2007a,fukuhara2007b}, Calcium \cite{kraft2009}, and Strontium \cite{martinez2009,stellmer2009} atomic gases have recently been obtained. In the present paper, we investigate theoretically the structure of diatomic molecules composed of one Strontium atom and one alkali atom (Li, Na, K, Rb, Cs), for the first time for most of these species. In the following, we will refer to them as ASr molecules. In particular, we determined the permanent dipole moment of their ground state in order to provide guidance to the experimentalists to choose the appropriate combination of atomic species. In contrast with alkali dimers, such species also have a magnetic moment in the ground state due to their three valence electrons, which therefore makes them suitable for their manipulation with both electric and magnetic external fields. In Section \ref{sec:method}, we first recall our methodology based on quantum chemistry computations involving Effective Core Potentials (ECP), reducing the problem to an effective three valence electron system, for which Full Configuration Interaction can be achieved. Molecular potential curves and permanent dipole moments for the LiSr, NaSr, KSr, RbSr, and CsSr molecules in their electronic ground state are presented in Sections \ref{sec:PEC} and \ref{sec:PDM}, emphasizing on the dipolar character of these molecules compared to bialkali molecules.


\section{Method of calculation} 
\label{sec:method}

We use the Configuration Interaction by Perturbation of a multi-configuration wave function Selected Iteratively (CIPSI) set of programs developed at the \textit{Université Paul Sabatier} in Toulouse (France). The CIPSI package incorporates both the use of Effective Core Potentials (ECP)\cite{durand1974,durand1975}, and $\ell$-dependent Core Polarization Potentials (CPP) \cite{muller1984a,foucrault1992}. Effective core potentials model the effect of inner shells electrons into a semi-empirical potential thus reducing the number of active electrons on each atom, namely one for the alkali atom, and two for the Strontium atom. Core polarization potentials embody the polarization effects of these effective cores with the valence ones. Thanks to the reduced number of active electrons, it becomes possible to perform a Full Configuration Interaction (FCI) in the valence space. This gives us a number of excited molecular states in addition to the lowest state in each symmetry. This method has proven to be efficient and accurate for alkali diatomics which are treated as two-electrons system \cite{aymar2005,aymar2007}, and we modified the CIPSI package accordingly for effective three-electron systems.

Effective core potentials are from ref.~\cite{fuentealba1982}. For the strontium atom, we use the so-called “large core” ECP for the 36 electrons of the Sr$^{2+}$ core, and basis sets composed of uncontracted Gaussian functions. The basis sets for alkali atoms are taken from ref.\cite{aymar2005}, where $f$ gaussian have been removed to reduce the size of the configuration space. We checked on the alkali atoms that this has no influence on the lowest energy levels we are interested in here. The basis set for the strontium atom is taken from~\cite{boutassetta1996}. We actually used two slightly different basis sets: we first designed sets for ASr$^+$ ions, \textit{i.e.} $8s7p4d$ for Lithium, $7s6p5d$ for Sodium, $7s5p7d$ for Potassium, $7s4p5d$ for Rubidium, $7s4p5d$ for Cesium, and $7s6p7d$ for Strontium ion whose exponents are reported in Table~\ref{tab:basisExpo}. These results will be discussed in a separate publication. For the neutral systems ASr, we reduced the Sr$^+$ set by a few orbitals down to $5s5p6d$, due to size constraints. The influence on the atomic energy levels remains small as it can be seen below.

In total, the number of configurations for the diatomic calculations is larger than 10$^5$, and depends on the molecular symmetries. We use a recently modified version of the $\ell$-dependent core polarization potential initially proposed by Foucrault \textit{et al.} \cite{foucrault1992}, which yields the correct asymptotic behavior of the potential energy at large internuclear distance \cite{guerout2010}, in contrast with most of the previous calculations performed on alkali dimers with this method. A set of cutoff radii, reported in Table~\ref{tab:cutoff}, are optimized to reproduce experimental values for the lowest $s$, $p$, and $d$ atomic levels of the alkali atoms and of the strontium cation (Table \ref{tab:Sr+levels}). Polarizabilities for the relevant cores Li$^{+}$, Na$^{+}$, K$^{+}$, Rb$^{+}$, Cs$^{+}$,and Sr$^{2+}$ cores are taken from experimental and semi-empirical values of refs.\cite{wilson1970,coker1976}.

\begin{table}
  \begin{center}
  \begin{tabular}{|c|c|c|c|c|c|c|}
    \hline
    angular momentum & Li & Na & K & Rb & Cs & Sr$^+$\\  \hline
    s & 2.464    &2.8357  & 0.9312  &1.292561&0.328926& 0.79174    \\
      & 1.991    &0.49318 & 0.2676  &0.824992&0.241529& 0.316178   \\
      & 0.582    &0.072085& 0.0427  &0.234682&0.050502& 0.066565   \\
      & 0.200    &0.036061& 0.03915 &0.032072&0.029302& 0.02699    \\
      & 0.070    &0.016674& 0.01448 &0.013962&0.013282& 0.013495   \\
      & 0.031    &0.00693 & 0.0055  &0.00575 &0.00528 & \textit{0.009}           \\
      & 0.015    &0.00287 & 0.0026  &0.0025  &0.003   & \textit{0.003}           \\
      & 0.007    &        &         &        &        &            \\ \hline
    p & 0.630    &0.431   & 0.133   &0.128   &0.1     & 0.225825   \\
      & 0.240    &0.09276 & 0.05128 &0.040097&0.0405  & 0.095691   \\
      & 0.098    &0.03562 & 0.01642 &0.014261&0.0162  & 0.042077   \\
      & 0.043    &0.01447 & 0.0052  &0.00485 &0.00443 & 0.018077   \\
      & 0.020    &0.0058  & 0.0022  &        &        & 0.0090385  \\
      & 0.010    &0.00023 &         &        &        & \textit{0.005}   \\
      & 0.005    &        &         &        &        &    \\ \hline
    d & 0.180    &0.292   & 1.255   &0.408807&0.196894& 3.618081   \\
      & 0.080    &0.06361 & 0.343   &0.096036&0.067471& 0.999656   \\
      & 0.22796  &0.02273 & 0.109   &0.026807&0.027948& 0.390735   \\
      & 0.008574 &0.008852& 0.0294  &0.009551&0.010712& 0.12277    \\
      &          &0.00352 & 0.01013 &0.004   &0.003   & 0.036655   \\
      &          &        & 0.0039  &        &        & 0.018327   \\
      &          &        & 0.0018  &        &        & \textit{0.009}            \\ \hline
symmetry                             & LiSr & NaSr & KSr  & RbSr & CsSr &  \\ \hline
$\{\Sigma^{+},\Delta_{x^{2}-y^{2}}\}$&129945&132792&161703&110337&110337&  \\
$\Pi$                                &120032&123684&151478&102540&102540&  \\
$\{\Sigma^{-},\Delta_{xy}\}$         &110240&114840&142138&95457 &95457 &  \\ \hline
  \end{tabular}
  \end{center}
  \caption{Gaussian basis set exponents used to represent the effective one-electron systems of the present work, and resulting number of configurations representing the effective three-electron diatomic systems A-Sr (with A $\equiv$ Li, Na, K, Rb, Cs) for the relevant molecular symmetries. The actual basis sets used for ASr neutral molecules include all exponents but the ones in italic, while the whole set has been used for ASr$^+$ systems. The three lowest lines provide the size of the hamiltonian for the Full Configuration Interaction.}
  \label{tab:basisExpo}
\end{table}

\begin{table}
  \begin{center}
    \begin{tabular}{ccccccc}
      \hline
                 & Li & Na & K & Rb & Cs & Sr$^+$\\  \hline
      $\rho_{s}$ & 1.4519565 &1.450275 & 2.10816   &2.5365&2.85585& 2.13999 \\
      $\rho_{p}$ & 1.013     &1.6452591& 1.9722    &2.279 &2.544  & 2.183 \\
      $\rho_{d}$ & 0.6       &1.5      & 1.9405    &2.5217&2.89815& 1.706 \\ \hline
                 &Li$^+$&Na$^+$&K$^+$&Rb$^+$&Cs$^+$&Sr$^{2+}$\\  \hline
      $\alpha$   & 0.1997    &0.9987& 5.472     &9.245&16.33& 5.67 \\ \hline
    \end{tabular}
  \end{center}
  \caption{Optimized cutoff radii $\rho_{\ell}$ (in atomic units a$_{0}$) and core polarizabilities $\alpha$ of the relevant ionic cores Li$^+$, Na$^+$, K$^+$, Rb$^+$, Cs$^+$ \cite{wilson1970}, Sr$^{2+}$ \cite{coker1976} entering into the Core Polarization Potentials (CPP).}
  \label{tab:cutoff}
\end{table}

Table \ref{tab:Sr+levels} shows that such an optimization yields a satisfactory representation of the Strontium cation spectrum. The corresponding fitted levels of the alkali atoms are not reported here, as we are mainly interested into the lowest level of $s$, $p$, and $d$ symmetry whose energies are represented exactly after the tuning of the cut-off radii of Table \ref{tab:cutoff}. Therefore the energies of a pair composed by one alkali atom and a Strontium cation in their lowest levels at infinite separations are well modeled here.


\section{The ground state potential curve of alkali-Strontium molecules} 
\label{sec:PEC}

As dissociation limits of ASr molecules involve a neutral alkali atom and a neutral Strontium atom, we first performed a first FCI calculation with the small and large basis sets and involving two electrons, in order to calculate the lowest strontium energy levels. Results are given in Table \ref{tab:Sr_levels}, and are compared to experimental values. The differences range between less than 200~cm$^{-1}$ for the ground state up to about 450~cm$^{-1}$ for the lowest excited levels, as it was the case also for the Magnesium atom \cite{aymar2009a}. We note that such an accuracy is typical for this kind of calculations, as illustrated also by ECP+CPP FCI calculations on Calcium \cite{gaveau2009}. The use of the small basis set does not significantly affect the quality of the results.

\begin{table}
\center
\begin{tabular} {|c|c|c|c|c|c|c|c|c|c|} \hline
State&Exp \cite{moore1952}&(a)&$\Delta_E$&(b)&$\Delta_E$&\cite{mitroy2008}&$\Delta_E$& \cite{guet1991}& $\Delta_E$   \\ \hline
 5s&-0.4053569&-0.405354&0   &-0.405358&0   &-0.405522&36  &-0.40839 &666\\
 4d&-0.3382685&-0.338269&0   &-0.338269&0   &-0.338247&-4  &-0.341908&799\\
 5p&-0.2948674&-0.294868&0   &-0.294869&0   &-0.294854&-2  &-0.29588 &222\\
 6s&-0.1878519&-0.187521&-73 &-0.187544&-68 &-0.187469&-84 &         &   \\
 5d&-0.1623282&-0.160856&-323&-0.160947&-303&-0.161237&-239&         &   \\
 6p&-0.1503744&-0.150234&-31 &-0.150237&-30 &-0.150150&-49 &         &   \\
 7s&-0.1093568&         &    &-0.109138&-48 &         &    &         &   \\
 6d&-0.9758806&         &    &-0.096543&-230&         &    &         &   \\
 7p&-0.921336 &         &    &-0.091920&-46 &         &    &         &   \\
 8s&-0.7166203&         &    &-0.075047&-34 &         &    &         &   \\ \hline
\end{tabular}
\caption {Binding energies (in a.u.) of Sr$^+$ levels obtained with (a) the small basis set, and (b) the large basis set for Sr$^+$ (Table \ref{tab:basisExpo}), and their deviations $\Delta_E$ (in cm$^{-1}$) from experiment \cite{moore1952}. The ionization limit of Sr$^+$ is taken at 88965.18~cm$^{-1}$ \cite{lange1991} to calculate the experimental binding energies in a.u.. We use the mass-corrected Rydberg constant 109736.63~cm $^{-1}$. The theoretical values of \cite{guet1991} have been spin-averaged.}
\label{tab:Sr+levels}
\end {table}

\begin{table}
  \begin{center}
    \begin{tabular}{|c||c|c|c|c|c||c|c|c|c|c|}    \hline
             &Exp&\multicolumn{2}{|c|}{(a)}&\multicolumn{2}{|c||}{(b)}&Exp&\multicolumn{2}{|c|}{(a)}&\multicolumn{2}{|c|}{(b)}\\ 
             &$E^{exp}_{b}$&$E_{b}$&$\delta_b$&$E_{b}$&$\delta_b$&$E^{exp}_{e}$&$E_{e}$&$\delta_e$&$E_{e}$&$\delta_e$ \\ \hline
$5s^2\,^1S$  &-0.61464148&-0.615471&-182&-0.615484&-184&0      &0    &0   &0    &0   \\
$5s5p\,^3P^o$&-0.54765155&-0.548880&-269&-0.548899&-273&14702.6&14613&-87 &14613&-88 \\
$5s4d\,^3D^e$&-0.53147099&-0.529868& 351&-0.529907& 343&18253.7&18779& 533&18781&528 \\
$5s4d\,^1D^e$&-0.52283208&-0.520684& 471&-0.520733& 460&20149.7&20793& 653&20795&645 \\
$5s5p\,^1P^o$&-0.51577530&-0.517756&-434&-0.517792&-442&21698.5&21447&-252&21440&-257\\ \hline
    \end{tabular}
  \end{center}
  \caption{Computed Strontium binding energies $E_{b}$ (in a.u. relative to the double ionization threshold of Strontium) and excitation energies $E_{e}$ (in cm$^{-1}$) compared to the experimental values of ref.\cite{moore1952} ($E^{exp}_{b}$ and $E^{exp}_{e}$), and their differences ($\delta_b=E_{b}-E^{exp}_{b}$ and $\delta_e=E_{e}-E^{exp}_{e}$) in cm$^{-1}$, resulting from a two-electron FCI calculation with (a) the large basis set, and (b) the small basis set for Sr$^+$ (Table \ref{tab:basisExpo}). The experimental values are averaged over fine structure energies when appropriate. We use the mass-corrected Rydberg constant 109736.63~cm $^{-1}$.}
  \label{tab:Sr_levels}
\end{table}

\begin{figure}[htb]
  \begin{center}
    \includegraphics[width=0.6\textwidth]{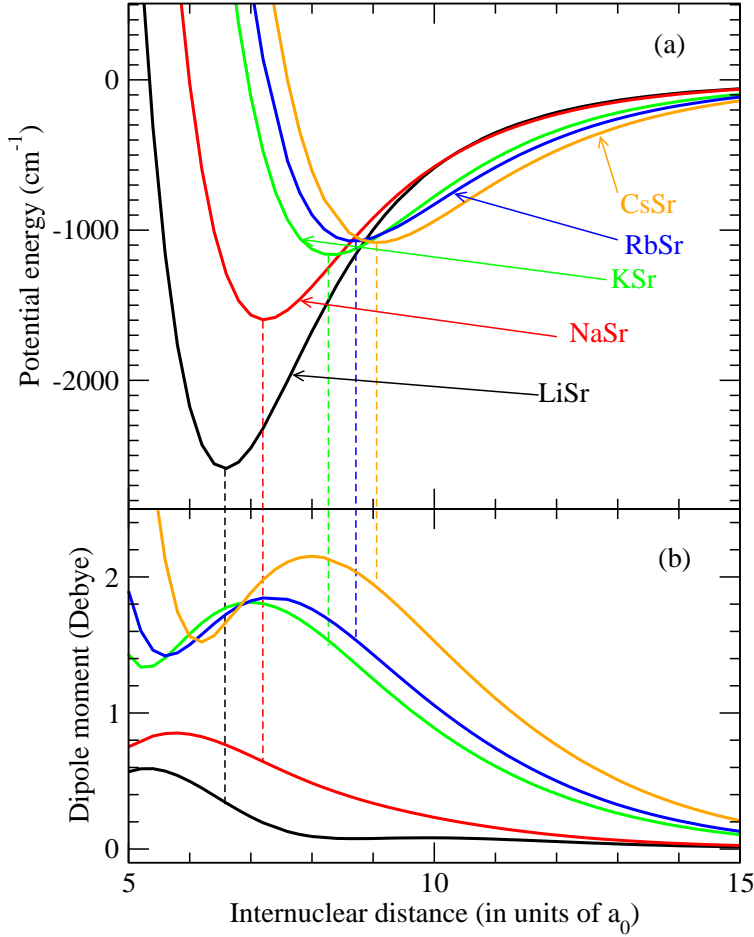}
  \end{center}
  \caption{(a) Potential energy curves and (b) permanent dipole moments (PDM) for the $X^2\Sigma^{+}$ ground state of the ASr molecules (A=Li, Na, K, Rb, Sr). Vertical lines guide the eye to locate the value of the PDM at the equilibrium distance.}
  \label{fig:ASr_doublet}
\end{figure}

We display in Figure \ref{fig:ASr_doublet}(a) the potential energy curves (PEC) for the $X^2\Sigma^+$ ground state of the ASr systems, relative to the same origin of energies fixed at the dissociation limit A$ns$+Sr($5s^2\,^1S$) for comparison purpose. The corresponding main spectroscopic constants are reported in Table \ref{tab:spectro}. The Strontium atom in its ground state is a closed-shell atom, so that the alkali-strontium molecules are expected to have a van der Waals character with a somewhat large equilibrium distance and weak binding energy, compared for instance to alkali diatomics with comparable masses like alkali-rubidium molecules. This feature is clearly visible in the figure, as the typical binding energies are three to five times smaller than for the related A-Rb molecules. The equilibrium distances are also systematically larger for ASr than for ARb molecules, except for LiSr. The LiSr molecule is a bit peculiar, as its ground state is quite deep, with an equilibrium distance comparable to the LiRb one.

The potential depths of the ASr molecules immediately show that in ultracold conditions these molecules will not be stable against collisions between them, just like KRb molecules \cite{ni2010}. The reaction channel yielding a Sr$_2$ molecule and an alkali dimer is energetically open for all alkali species. Indeed, the Sr$_2$ well depth is about 1061~cm$^{-1}$ \cite{stein2010}, while those of the alkali dimers are 8517~cm$^{-1}$ \cite{linton1996}, 6022~cm$^{-1}$ \cite{barrow1984}, 4451~cm$^{-1}$ \cite{amiot1991}, 3993~cm$^{-1}$ \cite{seto2000}, 3649~cm$^{-1}$ \cite{amiot2002a}, for Li$_2$, Na$_2$, K$_2$, Rb$_2$, Cs$_2$, respectively.

From the long range part of these curves we can estimate the coefficient $C_6$ of the leading term of the van der Waals interaction

\begin{table}
  \begin{center}
    \begin{tabular}{|c|c|c|c|c|c|c|}    \hline
    &$D_e$ (cm$^{-1}$)&$R_e$ (a.u.)& $\omega_e$ (cm$^{-1}$)&$B_e$  (cm$^{-1}$)& $d_e$ (Debye)&$d_0$ (Debye)\\ \hline
LiSr&2587&6.57&184.9&0.21&0.34&0.34 \\
NaSr&1597&7.22&85.4&0.063&0.62&0.63 \\
KSr &1166&8.33&52.4&0.032&1.52&1.50 \\
RbSr&1073&8.69&32.3&0.018&1.54&1.53 \\
CsSr&1084&9.06&33.8&0.013&1.91&1.91 \\ \hline
LiRb&13759&6.49&210&&4.42& \\
NaRb&5076&6.82&107&&3.41& \\
KRb &4199&7.63&76&&0.62& \\
CsRb&3907&8.29&&1.28&& \\ \hline
    \end{tabular}
  \end{center}
  \caption{Main spectroscopic constants for the lowest electronic states of the ASr molecules (A=Li, Na, K, Rb, Sr): well depth $D_e$, equilibrium distance $R_e$, harmonic constant $\omega_e$, and rotational constant $B_e$. The value of the dipole moment $d_e$ at the equilibrium distance, and averaged for the $v=0$ level ($d_0$) are also reported. The well depth, equilibrium distance, and permanent dipole moment at equilibrium of the related ARb molecules are displayed for comparison, following our computations reported in Refs. \cite{aymar2005,aymar2006,aymar2007}.}
  \label{tab:spectro}
\end{table}

\section{The permanent dipole moment of alkali-Strontium molecules} 
\label{sec:PDM}

As stated in the introduction, an important property of these molecules, in the perspective of cooling and trapping them at ultracold temperatures is the permanent dipole moment (PDM), which are drawn in Figure \ref{fig:ASr_doublet}(b) and reported as well in Table \ref{tab:spectro}. The PDM for ground state ASr molecules is predicted to be noticeable, even if they do not reach such high values than those of the alkali dimers \cite{aymar2005}. It is worthwhile to mention that the relation between the magnitude of the PDM and the difference in the mass (and then in the size) of the constituting atoms is inverted compared to alkali dimers: the PDM of LiCs and LiRb are the largest ones among all alkali pairs, while the PDM of LiSr is by far the smallest one compared to the other ASr species. This feature is amplified by the mismatch of the minimum of the PEC of LiSr and NaSr, and the maximum of the PDM curve, in contrast with the other species.

\section{Conclusion}

In this paper we investigate for the first time the electronic structure of the alkali-Strontium molecules with a quantum chemistry approach which accurately treat correlation between valence electron through a Full Configuration Interaction, while the correlation between valence and core electrons is accounted for through effective potentials. This approach has proven to be powerful for similar calculations on alkali dimers. The potential curve and the main spectroscopic constants, as well as the permanent dipole moment (PDM) of the $^2\Sigma^+$ ground state have been determined for LiSr, NaSr, KSr, RbSr, and CsSr molecules.  Though their PDM is generally smaller than the alkali-Rb molecules, these systems are possible candidates for achieving a new kind of ultracold molecular sample with anisotropic interactions. Just like heteronuclear alkali dimers, ultracold molecules in their ground state can probably be created by photoassociation, and this will be the topic of a further study.

While completing this paper, \.Zuchowski, Aldegunde and Hutson informed us about their study about the structure of the RbSr ground state \cite{zuchowski2010}, using an all-electron quantum chemistry approach. Their results are in good agreement with ours. For instance, they obtained an equilibrium distance $R_e=8.86$~a.u., a well depth of about 1000~cm$^{-1}$, and a PDM of 1.36~D at $R_e$ (compared to our values  $R_e=8.69$~a.u., $D_e=1073$~cm$^{-1}$, and $d_e=1.54$~D). The slight differences are probably due to the sensitivity of the results on the treatment of electron correlation, and further experiment should help to clarify the situation. In their paper, \.Zuchowski \textit{et al.} demonstrated the existence of Feshbach resonances in RbSr. Therefore, our next study on excited states of these molecules should guide experimentalists to design ways for achieving adiabatic population transfer from molecules created by Feshbach association down to the lowest energy level of the molecule, just like it has been recently done for Cs$_2$ \cite{danzl2008,danzl2010}and KRb \cite{ni2008} molecules.

\section*{Acknowledgments}
R.G. gratefully acknowledges support from \textit{Institut de Recherches sur les Atomes Froids} (IFRAF). This work has been performed in the framework of the network "Quantum Gases of Dipolar Molecules" (QuDipMol) of the EUROQUAM program of the ESF.

\section*{References} 
\label{sec:references}

\begin{thebibliography}{10}

\bibitem{baranov2007}
M.~A. Baranov,
\newblock Phys. Rep., {\bf 464},\hspace{0.25em}71  (2007).

\bibitem{anderson1995}
M.H. Anderson, J.R. Ensher, M.R. Matthews, C.E. Wieman, and E.A. Cornell,
\newblock Science, {\bf 269},\hspace{0.25em}198--201  (1995).

\bibitem{davis1995}
K.~B. Davis, M.~O. Mewes, M.~R. Andrews, N.~J. van Druten, D.~S. Durfee, D.~M.
  Kurn, and W.~Ketterle,
\newblock Phys. Rev. Lett., {\bf 75},\hspace{0.25em}3969  (1995).

\bibitem{bradley1995}
C.C. Bradley, C.A. Sackett, J.J. Tollett, and R.G. Hulet,
\newblock Phys. Rev. Lett., {\bf 75},\hspace{0.25em}1687  (1995).

\bibitem{griesmaier2005}
Axel Griesmaier, J\"org Werner, Sven Hensler, J\"urgen Stuhler, and Tilman
  Pfau,
\newblock Phys. Rev. Lett., {\bf 94},\hspace{0.25em}160401  (2005).

\bibitem{beaufils2008}
Q.~Beaufils, R.~Chicireanu, T.~Zanon, B.~Laburthe-Tolra, E.~Mar\'echal,
  L.~Vernac, J.-C. Keller, and O.~Gorceix,
\newblock Phys. Rev. A, {\bf 77},\hspace{0.25em}061601  (2008).

\bibitem{stuhler2005}
J.~Stuhler, A.~Griesmaier, T.~Koch, M.~Fattori, T.~Pfau, S.~Giovanazzi,
  P.~Pedri, and L.~Santos,
\newblock Phys. Rev. Lett., {\bf 95},\hspace{0.25em}150406  (2005).

\bibitem{beaufils2009}
Q.~Beaufils, A.~Crubellier, T.~Zanon, B.~Laburthe-Tolra, E.~Mar\'echal,
  L.~Vernac, and O.~Gorceix,
\newblock Phys. Rev. A, {\bf 79},\hspace{0.25em}032706  (2009).

\bibitem{aymar2005}
M.~Aymar and O.~Dulieu,
  sets,
\newblock J. Chem. Phys., {\bf 122},\hspace{0.25em}204302  (2005).

\bibitem{mancini2004}
M.W. Mancini, G.D. Telles, A.R.L. Caires, V.S. Bagnato, and L.G. Marcassa,
\newblock Phys. Rev. Lett., {\bf 92},\hspace{0.25em}133203  (2004).

\bibitem{wang2004}
D.~Wang, J.~Qi, M.~F. Stone, O.~Nikolayeva, B.~Hattaway, S.~D. Gensemer,
  H.~Wang, W.~T. Zemke, P.~L. Gould, E.~E. Eyler, and W.~C. Stwalley,
\newblock Eur. Phys. J. D, {\bf 31},\hspace{0.25em}165  (2004).

\bibitem{deiglmayr2008}
J.~Deiglmayr, M.~Aymar, R.~Wester, M.~Weidem\"uller, and O.~Dulieu,
\newblock J. Chem. Phys., {\bf 129},\hspace{0.25em}064309  (2008).

\bibitem{haimberger2004}
C.~Haimberger, J.~Kleinert, M.~Bhattacharya, and N.P. Bigelow,
\newblock Phys. Rev. A, {\bf 70},\hspace{0.25em}21402  (2004).

\bibitem{kerman2004}
A.~J. Kerman, J.~M. Sage, S.~Sainis, T.~Bergeman, and D.~DeMille,
\newblock Phys. Rev. Lett., {\bf 92},\hspace{0.25em}153001  (2004).

\bibitem{ni2008}
K.-K. Ni, S.~Ospelkaus, M.~H.~G. de~Miranda, A.~Peer, B.~Neyenhuis, J.~J.
  Zirbel, S.~Kotochigova, P.~S. Julienne, D.~S. Jin, and J.~Ye,
\newblock Science, {\bf 322},\hspace{0.25em}231  (2008).

\bibitem{ospelkaus2010b}
G.~Qu\'em\'ener B. Neyenhuis D. Wang M.H.G. deMiranda J.L. Bohn J.~Ye
  S.~Ospelkaus, K-K~Ni and D.S. Jin,
\newblock Phys. Rev. Lett., {\bf 104},\hspace{0.25em}030402  (2010).

\bibitem{danzl2010}
J.~G. Danzl, M.~J. Mark, E.~Haller, M.~Gustavsson, R.~Hart, J.~Aldegunde, J.~M.
  Hutson, and H.-C. N\"agerl,
\newblock Nature Phys., {\bf 6},\hspace{0.25em}265  (2010).

\bibitem{meyer2009}
E.~R. Meyer and J.~L. Bohn,
\newblock Phys. Rev. A, {\bf 80},\hspace{0.25em}042508  (2009).

\bibitem{takasu2004}
Y.~Takasu, K.~Komori, K.~Honda, M.~Kumakura, T.~Yabuzaki, and Y.~Takahashi,
\newblock Phys. Rev. Lett., {\bf 93},\hspace{0.25em}123202  (2004).

\bibitem{kitagawa2008}
M.~Kitagawa, K.~Enomoto, K.~Kasa, Y.~Takahashi, R.~Ciurylo, P.~Naidon, and
  P.~S. Julienne,
\newblock Phys. Rev. A, {\bf 77},\hspace{0.25em}012719  (2008).

\bibitem{nemitz2010}
N.~Nemitz, F.~Baumer, F.~M\"{u}nchow, S.~Tassy, and A.~G\"orlitz,
\newblock Phys. Rev. A, {\bf 79},\hspace{0.25em}061403  (2010).

\bibitem{fukuhara2007a}
T.~Fukuhara, S.~Sugawa, and Y.~Takahashi,
\newblock Phys. Rev. A, {\bf 76},\hspace{0.25em}051604  (2007).

\bibitem{fukuhara2007b}
T.~Fukuhara, Y.~Takasu, M.~Kumakura, and Y.~Takahashi,
\newblock Phys. Rev. Lett., {\bf 98},\hspace{0.25em}030401  (2007).

\bibitem{kraft2009}
S.~Kraft, F.~Vogt, O.~Appel, F.~Riehle, and U.~Sterr,
\newblock Phys. Rev. Lett., {\bf 103},\hspace{0.25em}130401  (2009).

\bibitem{martinez2009}
Y.~N.~Martinez de~Escobar, P.~G. Mickelson, M.~Yan, B.~J. DeSalvo, S.~B. Nagel,
  and T.~C. Killian,
\newblock Phys. Rev. Lett., {\bf 103},\hspace{0.25em}200402  (2009).

\bibitem{stellmer2009}
S.~Stellmer, M.-K. Tey, B.~Huang, R.~Grimm F., and Schreck,
\newblock Phys. Rev. Lett., {\bf 103},\hspace{0.25em}200401  (2009).

\bibitem{durand1974}
P.~Durand and J.C. Barthelat,
\newblock Chem. Phys. Lett., {\bf 27},\hspace{0.25em}191  (1974).

\bibitem{durand1975}
P.~Durand and J.C. Barthelat,
\newblock Theor. chim. Acta, {\bf 38},\hspace{0.25em}283  (1975).

\bibitem{muller1984a}
W.~M\"uller, J.~Flesch, and W.~Meyer,
\newblock J. Chem. Phys., {\bf 80},\hspace{0.25em}3297  (1984).

\bibitem{foucrault1992}
M.~Foucrault, Ph. Milli\'e, and J.P. Daudey,
\newblock J. Chem. Phys., {\bf 96}(2),\hspace{0.25em}1257  (1992).

\bibitem{aymar2007}
M.~Aymar and O.~Dulieu,
\newblock Mol. Phys., {\bf 105},\hspace{0.25em}1733  (2007).

\bibitem{fuentealba1982}
P.~Fuentealba, H.~Preuss, H.~Stoll, and L.~Von Szentpaly,
\newblock Chem. Phys. Lett., {\bf 89},\hspace{0.25em}418  (1982).

\bibitem{boutassetta1996}
N.~Boutassetta, A.~R. Allouche, and M.~Aubert-Fr\'econ,
\newblock Phys. Rev. A, {\bf 53},\hspace{0.25em}3845  (1996).

\bibitem{guerout2010}
R.~Gu\'erout, O.~Dulieu, and F.~Spiegelman,
\newblock in preparation.

\bibitem{wilson1970}
J.~N. Wilson and R.~M. Curtis,
\newblock J. Chem. Phys., {\bf 74},\hspace{0.25em}187  (1970).

\bibitem{coker1976}
H.~Coker,
\newblock J. Phys. Chem., {\bf 80},\hspace{0.25em}2073  (1976).

\bibitem{aymar2009a}
M.~Aymar, R.~Gu\'erout nd~M.~Sahlaoui, and O.~Dulieu,
\newblock J. Phys. B, {\bf 42},\hspace{0.25em}154025  (2009).

\bibitem{gaveau2009}
M.-A. Gaveau, J.-M. Mestdagh, T.~Bouissou, G.~Durand, M.-C. Heitz, and
  F.~Spiegelman,
\newblock Chem. Phys. Lett., {\bf 467},\hspace{0.25em}260  (2009).

\bibitem{moore1952}
C.E. Moore,
\newblock {\em Atomic energy levels, vol. 2 (Chromium through Niobium)},
\newblock (U.S. government printing office, Washington,  1952).

\bibitem{mitroy2008}
J.~Mitroy and J.-Y. Zhang,
\newblock J. Chem. Phys., {\bf 128},\hspace{0.25em}134305  (2008).

\bibitem{guet1991}
C.~Guet and W.~R. Johnson,
\newblock Phys. Rev. A, {\bf 44},\hspace{0.25em}1531  (1991).

\bibitem{lange1991}
V.~Lange, M.~A. Khan, U.~Eichmann, and W.~Sandner,
\newblock Zeit. Phys. D, {\bf 18},\hspace{0.25em}319  (1991).

\bibitem{ni2010}
D.~Wang G. Qu\'em\'ener B. Neyenhuis M. H. G. de Miranda J. L. Bohn J. Ye D.
  S.~Jin K.-K.~Ni, S.~Ospelkaus,
\newblock Nature, {\bf 464},\hspace{0.25em}1324  (2010).

\bibitem{stein2010}
A.~Stein, H.~Kn\"ockel, and E.~Tiemann,
\newblock Eur. Phys. J. D, {\bf 57},\hspace{0.25em}171  (2010).

\bibitem{linton1996}
C.~Linton, F.~Martin, I.~Russier, A.~J. Ross, P.~Crozet, S.~Churassy, and
  R.~Bacis,
\newblock J. Mol. Spectrosc., {\bf 175},\hspace{0.25em}340  (1996).

\bibitem{barrow1984}
R.~F. Barrow, J.~Verg\`es, C.~Effantin, K.~Hussein, and J.~d'Incan,
\newblock Chem. Phys. Lett., {\bf 104},\hspace{0.25em}179  (1984).

\bibitem{amiot1991}
C.~Amiot,
\newblock J. Mol. Spectrosc., {\bf 147},\hspace{0.25em}370  (1991).

\bibitem{seto2000}
J.~Y. Seto, R. J.~Le Roy, J. Verg\`es, and C. Amiot,
\newblock J. Chem. Phys., {\bf 113},\hspace{0.25em}3067--3076
  (2000).

\bibitem{amiot2002a}
C.~Amiot and O.~Dulieu,
\newblock J. Chem. Phys., {\bf 117},\hspace{0.25em}5155  (2002).

\bibitem{aymar2006}
M.~Aymar, O.~Dulieu, and F.~Spiegelman,
\newblock J. Phys. B: At. Mol. Opt. Phys., {\bf 39},\hspace{0.25em}S905
  (2006).

\bibitem{zuchowski2010}
P.~S. \.Zuchowski, J.~Aldegunde, and J.~M. Hutson,
\newblock arXiv:1006.3006v1 [physics.atom-ph]  (2010).

\bibitem{danzl2008}
J.~G. Danzl, E.~Haller, M.~Gustavsson, M.~J. Mark, R.~Hart, N.~Bouloufa,
  O.~Dulieu, H.~Ritsch, and H.-C. N\"{a}gerl,
\newblock Science, {\bf 321},\hspace{0.25em}1062  (2008).

\end{thebibliography}

\end{document}